# Parallel nonlinear neuromorphic computing with temporal encoding

Guangfeng You[1,2], Chao Qian[1,2,*], and Hongsheng Chen[1,2,*]

[1] *ZJU-UIUC Institute, Interdisciplinary Center for Quantum Information, State Key Laboratory of Extreme Photonics and Instrumentation, Zhejiang University, Hangzhou 310027, China.*

[2] *ZJU-Hangzhou Global Science and Technology Innovation Center, Zhejiang Key Laboratory of Intelligent Electromagnetic Control and Advanced Electronic Integration, Zhejiang University, Hangzhou 310027, China.*

[*]*Corresponding authors:* chaoq@intl.zju.edu.cn *(C. Qian);* hansomchen@zju.edu.cn *(H. Chen)*

**Abstract:** The proliferation of deep learning applications has intensified the demand for electronic hardware with low energy consumption and fast computing speed. Neuromorphic photonics have emerged as a viable alternative to directly process high-throughput information at the physical space. However, the simultaneous attainment of high linear and nonlinear expressivity posse a considerable challenge due to the power efficiency and impaired manipulability in conventional nonlinear materials and optoelectronic conversion. Here we introduce a parallel nonlinear neuromorphic processor that enables arbitrary superposition of information states in multi-dimensional channels, only by leveraging the temporal encoding of spatiotemporal metasurfaces to map the input data and trainable weights. The proposed temporal encoding nonlinearity is theoretically proved to flexibly customize the nonlinearity, while preserving quasi-static linear transformation capability within each time partition. We experimentally demonstrated the concept based on distributed spatiotemporal metasurfaces, showcasing robust performance in multi-label recognition and multi-task parallelism with asynchronous modulation. Remarkably, our nonlinear processor demonstrates dynamic memory capability in autonomous planning tasks and real-time responsiveness to canonical maze-solving problem. Our work opens up a flexible avenue for a variety of temporally-modulated neuromorphic processors tailored for complex scenarios.

**Introduction**

The ability of quickly processing the incessant influx of data is highly demanded for state-of-the-art artificial intelligence algorithms[1-3], especially amid escalating global reliance on smart devices[4] and augmented reality applications[5,6]. Graphics processing units and other application-specific electronic hardware accelerators have greatly augmented the performance in terms of computational speed and energy, however, the approach to physical limit in semiconductor technology motivates scientists to exploit new computing paradigm[7]. Optical neural network[8-10], an isomorphism between photonics structures and mathematical formalisms, has found to be a promising alternative to process information at the physical space, with the iconic features of exceptional computing speed and minimal energy consumption. In general, optical neural network encompasses both linear (e.g., vector-matrix multiplication)[11] and nonlinear operation (e.g., activation function)[12]. Among them, linear operation has been extensively investigated by controlling the scattering and diffraction behaviour of light to emulate multiple-input multiple-output matrix[13]. Nonlinear operation, albeit extremely challenging, plays an indispensable role for brain-like casual inference[14], empowering neural network to approximate arbitrary functions with adaptive connections among neurons[15]. For instance, by utilizing the phase encoding mechanism, facilitated by all-optical conjugation and phase-to-intensity transformation, exciting applications such as random wavefront distortion correction and spectral filter can be effectively achieved[16,17]. So far, nonlinear implementations predominately rely on optoelectronic conversion[18-21] and inherent material properties[22-23], such as saturated absorption and optical bistability[24,25]. Nevertheless, these methodologies remain great challenges associated with energy efficiency, controlling system, and computing latency.

A novel avenue towards optical nonlinear operation is to leverage unique information encoding strategies that induce nonlinear effects with linear scattering on the platform of reverberating cavity[26], multiple spatial light modulator[27], and racetrack resonators[28], and more. Replicas of information are encoded into reconfigurable linear photonic structures, with the higher-order states of modulated information being excited through multiple scattering, while maintaining the linear nature of the propagating wave. Such structural encoded nonlinearity empowers nonlinear computation within linear systems, which promotes a reinterpretation of equivalent optical nonlinearity[29]. However, the quest for the optimal data-encoding nonlinearity still remains challenging, primarily due to the fact that the trainable parameters are exclusively explored in spatial dimension, resulting in a finite-dimensional solution space. Moreover, there exist an upper limit to network performance because this form of spatial encoding nonlinearity ultimately manifests as a point spread function that depends on the spatial variation of the information data[30]. This intrinsic limitation gives rise to a fundamental trade-off between diminished linear performance and enhanced nonlinearity. Consequently, how to achieve high-performance nonlinear and linear manipulation in optical neural network still persists as an unresolved challenge.

Here we report a parallel nonlinear neuromorphic processor by leveraging the temporal sequence of spatiotemporal metasurfaces[31-33]. Through editing the temporal sequence, the meta-atoms/neurons on the metasurfaces are conceptualized as time-varying media, which facilitates independent linear and nonlinear manipulations through weight adjustment in discrete time partitions. The employed temporal encoding methodology mitigates mutual interference between linear and nonlinear effects, while its unique temporal multiplexing attributes furnish a robust hardware platform capable of supporting extensive task parallelism across multiple frequencies. In the experiment, multiple spatiotemporal metasurfaces are interconnected in a distributed manner to construct a nonlinear computing space, wherein the input data and weight matrices are represented and trained by the temporal sequence. Such temporal nonlinear neural network is demonstrated for multi-label recognition, multi-task parallelism with asynchronous modulation, and reinforcement learning agent in maze-solving problem. This temporally encoding induced nonlinearity expands the scope of existing neuromorphic computing and unlocks a plethora of exciting applications in optical information display[34,35] and encrypted communication[36].

**Theoretical analysis of nonlinear mapping by temporal encoding**

Our proposed nonlinear system is constructed by randomly-distributed spatiotemporal metasurfaces, each of which has $M \times N$ unit cells. We hypothesize that the spatiotemporal metasurfaces are controlled by periodic temporal sequences with the modulation period of $T_0$ ($L$ time partitions). The input data $D$ is modulated into a temporal sequence, whereby the nonlinearity between input data and system response is excited through encoding diffractions, as illustrated in Fig. 1. The interplay of multiple encoding diffractions among these metasurfaces modulates the scattering field according to the individual time-varying periodic sequence $S = [D, W_1, D, W_2, D, W_3, ...]$ attributed to each metasurface. Here $S$ is composed of the input data $D$ and multiple weight information $W$ (Methods). The high-order state of $D$ arises from the interaction caused by multiple scattering as data information propagates between metasurfaces. Specifically, assuming the incident field $A$ with the wavelength of $\lambda_c$, the corresponding propagation function at $(\xi, \eta, z)$ for an individual metasurface driven by independent data and weight information state[37] can be expressed as follows[38,39]:

$$F(\xi, \eta, t) = \frac{-i}{2\lambda} \iint dx dy \Gamma_{xy}(t) A(x,y) (1+\frac{z}{r}) \frac{e^{i\frac{(x-\xi)^2-(y-\eta)^2}{\lambda_c}}}{r} \quad (1)$$

where $\Gamma_{xy}(t)$ is the reflection coefficient of the meta-atom at $(x,y)$ and $\Gamma_{xy}(t) = \sum_l \Gamma_{xy}^l(t) U^l(t) = D_{xy} \cdot U^0(t) + W_{1,xy}^{T_0} \cdot U^1(t) + \ldots + D_{xy} \cdot U^{l-1}(t) + W_{l,xy}^{T_0} \cdot U^l(t) + \ldots + D_{xy} \cdot U^{L-1}(t) + W_{L,xy}^{T_0} \cdot U^L(t)$. Without loss of generality, $W_{l,xy}^{T_0}$ represents the weight matrix of meta-atom $(x,y)$ in the $l_{th}$ time partition of periodic temporal sequence. $U^l(t)$ is a unit step function at the $l_{th}$ time partition, which can be expressed in Fourier expansion $U^l(t) = [U(t - l \cdot \tau) - U(t - (l+1) \cdot \tau)] = \sum_{m=-\infty}^{\infty} e^{j2\pi m \Delta f_0} \cdot a_{lm}$. The propagation function $F_m$ at the harmonic $f_c + m\Delta f_0$ is

$$F_m(\xi, \eta) = \frac{-i}{2\lambda} \iint dx dy \sum_l a_{ml} (D_{xy} + e^{-i2\pi lm\Delta f_0 \tau} W_{l,xy}^{T_0}) A(x,y)(1+\frac{z}{r}) \frac{e^{i\frac{(x-\xi)^2-(y-\eta)^2}{\lambda_c}}}{r} \quad (2)$$

For a given modulation period $T_0$, considering the linear diffraction operator and constant $a_{ml}$, response of $l_{th}$ metasurface layer to the incidence $E_{in}$ for three trainable weight matrices $W^{T_0}$ can be expressed into a matrix form:

$$f_l(D, W^{T_0}) = E_{in}(a \cdot D + b \cdot W_1^{l,T_0} + c \cdot W_2^{l,T_0} + d \cdot W_3^{l,T_0}) \quad (3)$$

Although the relation between input data $D$ and propagation function is linear, the system response of multiple scattering is a nonlinear function about $D$. For example, three reflective metasurfaces are used to build a temporal nonlinear network and thus the system response can be expressed as follows:

$$y = E_{in} f_1(D, W^{1,T_0}) f_2(D, W^{2,T_0}) f_3(D, W^{3,T_0}) G = E_{in} G \prod_l f_l(D, W^{l,T_0}) \quad (4)$$

$G$ is a linear diffraction operator used to characterize the linear propagation of wave and $\prod$ is a product operator. This temporally encoding induced optical nonlinearity constitutes the physical foundation of our work and the order of nonlinearity is augmented by extending the length of time-varying sequences and the number of metasurfaces. Similar encoding nonlinearity can also be found in far-field propagation of spatiotemporal metasurfaces in Supplementary Note 1.

We would like to emphasize that our temporal encoding strategy enhances the nonlinearity of optical neural network without sacrificing its linear expressivity. State-of-the-art spatial data repetition modulates the information with a certain spatial encoding function[30], such as $h_D(x,y) = W \cdot D_{xy} + b$, thus converting a general linear diffraction mechanism into input-encoding-dependent point spread function $y = E_{in}G(W \cdot D_{xy} + b) = E_{in}Gh_D(x,y)$, which undermines the network capacity for linear transformation. However, in our work, instead of pre-processing the input data with scaling and bias parameters, both data and weight matrices are equally arranged in distinct time partitions without additional spatial encoding function $h$, as indicated in Eqs. (3) and (4). This consideration enables linear propagation between adjacent layers, while the whole system response exhibits nonlinearity, allowing us to achieve flexible linear and nonlinear performance.

**Experimental implementation of temporal nonlinear network**

Guided by the temporally encoding nonlinearity in Eq. (4), we present a simple yet feasible temporal nonlinear network composed of three randomly-distributed spatiotemporal metasurfaces, as displayed in Fig. 2. Each unit cell at metasurface layer behaves as a neuron of neural network, and the neuron $(x,y)$ is applied by a time-varying sequence encompassing the input data $D_{xy}$ and trainable elements $[W_{1,xy}, W_{2,xy}, ..., W_{L,xy}]$, of which the weight elements are iteratively updated during backward propagation. In the training process, as shown in Fig. 2a, $D$ and $W$ are temporally modulated in each metasurface layer, and the output results are detected by corresponding receivers. The detected results, combined with target outcomes, are utilized to calculate the weight gradient to minimize the loss function, i.e., corrected categorical cross-entropy[40]; see the detailed gradient derivation in Supplementary Note 2. To achieve optimal discrete optimization, a probability-based strategy is introduced to transform continuous probability distribution into discrete state distribution (weight matrices) in the updating process, where the selection of state representation is flexible and can map to the physical parameters, such as phase or amplitude. The discrete weight matrix is quantized into continuous statistical function, that is, each matrix element follows a Bernoulli distribution with probability $\rho$, which allows for updating discrete weights through updating $\rho$ with the soft-argmax function[41].

As demonstration in Fig. 2b, we perform a microwave experiment and observe the transformation from linear to nonlinear, which can also be readily generalized to optical frequencies by optical reconfigurable metasurfaces[42]. In the following experiment, the phase information is employed to differentiate various states of unit cells and the reflection spectra for reconfigurable metasurfaces embedded with PIN diode are depicted in Fig. 2c. Figure 2d demonstrates the experimental setup of nonlinear neural network architecture, where each spatiotemporal metasurface contains $12 \times 12$ unit cells. For a given task, the input information $D$ is modulated into periodic sequences $S$ of metasurfaces with field programmable gate array (FPGA) and the trained temporal nonlinear network processes the input information through physical diffractions, subsequently generating a corresponding output intensity distribution that can be detected by receivers.

**Multi-label recognition with temporal nonlinear network**

We experimentally benchmark our model in multi-label recognition tasks with facial images[43]. As displayed in Fig. 3a, the original images with $144 \times 144$ pixels are down-sampled into $12 \times 12$ binarized images and mapped to the metasurface one-by-one, which leads to a diminished input aperture (i.e., input field-of-view). Even though such data processing method yields limited dimensionality of the transformation solution space for linear diffraction processor[44], its impact on our network is minimal because the nonlinear expressivity enables the recovery of the original image features (Supplementary Note 3). Then, the binarized image $D$, combined with weight matrices $W$, are used to formulate the periodic temporal sequence $S = [D, W_1, D, W_2, D, W_3, D, W_4]$ for each metasurface during network training. Here the length of each temporal sequence is set to 8 and $T_0$ of the time-varying sequence is set as 8 $\mu s$, that is, each trainable layer of nonlinear temporal network contains four $12 \times 12$ reconfigurable weight matrices $[W_1, W_2, W_3, W_4]$. Figure 3b illustrates that the maximum energy concentration and intensity distribution of surface field at the second layer vary for different input image data $D$. This variation implies the propagation of data information flow among metasurfaces, which can be directly visualized in the attention map represented in the form of heat map[45]. It can be observed that the focal region of the attention map typically corresponds to those parts of the input image that contain key label information, consistent with the characteristics of Poisson propagation trajectories (Supplementary Note 4). As demonstrated in Fig. 3c, the received intensity represents the level of classification possibility and the receiver with the maximum receiving intensity corresponds to the classification result. More specifically, the variance of receiving intensity over its original intensity $\ell(E) = \Delta E/E_0$ is quantized as an indicator to represent the network output.

Entropy coefficient (EC) is also utilized to evaluate the recognition performance[46], whereby a smaller EC represents a better recognition accuracy and a greater intensity difference between receivers.

Since the neurons/unit cells of metasurface are controlled by temporal encoding strategy, the fully connected pointwise multiplication operation can be further generalized into a convolution process between fields-of-view (FOV) and weight filters[47], possessing strong compatibility with different input data size and format. As shown in Fig. 4a, the original images with $2M \times 2N$ pixels can be artificially divided into four data FOVs $\boldsymbol{D} = [\boldsymbol{D}_1, \boldsymbol{D}_2, \boldsymbol{D}_3, \boldsymbol{D}_4]$, each of which has $M \times N$ pixels. The FOVs and weight filters $\boldsymbol{W}$ are arranged within the spatial dimension, corresponding to the shifts of filters across the images, as well as within the temporal dimension, corresponding to different FOV and filters. The corresponding time-varying sequence at $i_{th}$ metasurface layer can be expressed as $\boldsymbol{S} = [\boldsymbol{D}_1, \boldsymbol{W}_1^i, \boldsymbol{D}_2, \boldsymbol{W}_2^i, \boldsymbol{D}_3, \boldsymbol{W}_3^i, \boldsymbol{D}_4, \boldsymbol{W}_4^i]$. Consequently, each neuron computes a summed weighted output $o$ as follows:

$$o = \sum_{x,y} \sum_l \sum_k D_{l,xy}^i W_{k,xy}^{i+1} + D_{k,xy}^{i+1} W_{l,xy}^i \quad (5)$$

where $D_{l,xy}^i$ and $W_{l,xy}^i$ are the image data and weight of unit cell $(x, y)$ on $l_{th}$ time partition at $i_{th}$ metasurface layer, respectively. Utilizing the data repetition strategy, three metasurface layers are concurrently manipulated through image data and weight matrices, resulting in an improved residual convolutional process as indicated in Eq. (5). Figure 4b provides the accuracy comparison for the cases with the image size of $24 \times 24$ and $48 \times 48$, where the enhanced recognition performance further illustrates the superiority of the inherent residual convolutional mechanism of temporal nonlinear network in image processing. Similar observations are also drawn in the confusion matrix in Fig. 4c.

**Nonlinear multi-task parallelism with asynchronous modulation**

Furthermore, our nonlinear computing system holds great potential for massively parallel computing by harnessing different modulation frequency to manipulate the temporal sequences of the distinct metasurface areas, known as asynchronous modulation[48,49]. By employing different modulation frequencies and time delay configurations, a given time segment can be split into temporal sequences with different lengths on multiple harmonics, respectively. This enhanced temporal multiplexing strategy enables independent manipulation of different tasks on these harmonics through precise setting of the time sequence durations. In the proof of concept, as shown in Fig. 5a, we perform two independent tasks simultaneously by dividing each metasurface of our nonlinear system into two individual frequency partitions, with partition 1 being modulated with frequency $\Delta f_1$ and partition 2 being modulated with $\Delta f_2$, respectively. Here the modulation frequency $\Delta f_1$ of metasurface time-varying sequence is set as 0.0625 MHz, while $\Delta f_2$ is set as 0.125 MHz. It should be note that no theoretical correlation exists between the modulation frequencies $\Delta f_1$ and $\Delta f_2$, though a multiplicative relation is observed in the following confirmatory experiment (i.e., $\Delta f_2 = 2\Delta f_1$). By reasonably adjusting the modulation frequency of time-varying sequence for different spatial partitions of metasurfaces, our improved asynchronous modulation technique is able to manipulate different harmonics independently. $\Delta f_1$ and $\Delta f_2$ frequency partitions are determined based on the sensitivity of the data and weight information states in time-varying sequences. As displayed in Fig. 5b, single-pixel information state sensitivity is introduced to quantify the sensitivity of the output intensity to the variation of the matrix elements at each time partition[50], which is defined here as the entropy difference for variations on each individual unit cell of metasurface. Then total information state sensitivity that characterizes the overall sensitivity of all pixels in single metasurface layer is derived from the union of the single-pixel information state sensitivities (Methods). Following the distribution of total information state sensitivity, different frequency manipulation partitions for each metasurface are finally determined, which are subsequently employed for asynchronous metasurface modulation. To achieve the parallelism in our asynchronous temporal nonlinear system, the high sensitivity area and low sensitivity area of the total information state sensitivity distribution are modulated with the frequency $\Delta f_1$ and $\Delta f_2$, respectively. We propose to reframe parallel computing as Lagrangian optimization problem and this objective can be optimized using modified wake-sleep algorithm[51], as shown in Fig. 5a. Four weight matrices are utilized for feature extraction at frequency $\Delta f_1$ and $\Delta f_2$ originally and additional trainable weight matrices are extended in the sequences at frequency $\Delta f_2$ for subsequent different tasks (Methods). Therefore, the meta-atoms in partition 1 at each metasurface layer are controlled by the time-varying sequence $S_{\Delta f_1} = [D, W_1^{\Delta f_1}, D, W_2^{\Delta f_1}, D, W_3^{\Delta f_1}, D, W_4^{\Delta f_1}]$, while the meta-atoms in partition 2 are controlled by $S_{\Delta f_2} = [D, W_1^{\Delta f_2}, D, W_2^{\Delta f_2}, D, W_3^{\Delta f_2}, D, W_4^{\Delta f_2}, D, W_5^{\Delta f_2}, D, W_6^{\Delta f_2}, D, W_7^{\Delta f_2}, D, W_8^{\Delta f_2}]$. In Fig. 5c, we tested the recognition performance on two different classification tasks at harmonics $f_c + 0 \times \Delta f_2$ and $f_c +$ $1 \times \Delta f_1$. For a given input image, our nonlinear system extracts the physical features at different frequencies tailored for specific task requirement. Effective recognition across multiple frequencies reveals the advantages of our asynchronous system in facilitating highly parallel neuromorphic computing and provides a novel hardware platform for high-efficiency large-scale parallelism.

**Reinforcement learning based maze-solving with temporally induced nonlinear agents**

To further evaluate our nonlinear physical system, we delve into complex scenarios to enable it as an intelligent nonlinear agent for maze-solving and reinforcement learning algorithm. As illustrated in Fig. 6a, the actual maze scenes are mapped into the input data $D$ to metasurfaces. Since each movement of the vehicle will lead to a slight change of metasurface distribution, the length of input sequence $S = [D, W_1, D, W_2, ..., D, W_{128}]$ is set to 256, of which the maze data $D$ is repeated by 128 times and the number of trainable weight matrices $W$ are chosen as 128 to capture small differences of maze data. In the training process, a number of maze data and corresponding reward values are stored in memory pool and randomly selected as inputs for each iteration. The temporal nonlinear neural network is employed to construct Q network and target Q' network for decision-making (Methods). For the random locations inside maze (Fig. 6b), the trained metasurface agent can give the optimal direction policy to get out of the maze. We measured the intensity at stable state for each detector and the maximum intensity appears at the expected receiving location, with the average variance of intensity significantly stronger than that of other receivers. The metasurface agent possesses memory capabilities during the exploration process, allowing a dynamic response to the real-time environment based on multiple timesteps of past events in memory pool. Specifically, for a vehicle at any location of the maze, the metasurface agent is capable of providing continuous policy manipulation and facilitates the vehicle's egress from the maze. As shown in the Fig. 6c, we show a complete process of maze-solving step by step from the entrance to the exit of the maze. It is worth noting that, in order to complete this continuous searching, the agent needs to be fully trained, because the direction policy errors at some key locations will lead to the failure of continuous searching; see more intermediate training results in Supplementary Note 5.

**Discussion**

In conclusion, we have proposed and experimentally demonstrated a multi-functional nonlinear metasurface processor for neuromorphic computing. Our unique nonlinear methodology elucidates a novel pathway for transforming linear scattering into nonlinear computational process, ingeniously maintaining linearity in the spatial modulation while achieving nonlinearity in the temporal domain. Through multi-dimensional encoding nonlinear implementation, we have realized a temporal nonlinear neural network that exhibits compatibility across other spectra and diverse scenarios without reliance on complex nonlinear materials or high-power excitation, which may inspire other applications, such as meta-hologram[52], neuromorphic neural circuits[53], and adaptive focusing localization[54]. Moving forward, the distinctive nature of information propagation stimulates us to re-examine conventional scattering theory through the lens of information flow that the optical scattering does not constitute an information-destroying process. Instead, our system provides a groundbreaking paradigm of information propagation and holds significant potential for wide applications in data storage[55], optical encryption[56], and reliable coherent communications[57].

## Methods

### Experimental setup and information state manipulation

The excitation source is a standard gain antenna from 3.94 GHz to 5.99 GHz, and four standard gain antennas with an axial distance of 33 mm from the third metasurface are used as receivers. Theses receivers are connected to vector network analyzer (VNA) via radio-frequency (RF) converter. The computer reads the measurement results of VNA via Ethernet and uses FPGA to feed back the updated modulation sequence to the metasurfaces. The experiment was conducted in a microwave anechoic chamber. Three reflective metasurfaces with the size of 230 mm × 230 mm are relative to each other, as shown in Fig. S1. Each metasurface has 12 × 12 reconfigurable unit cells, and the bias state of each unit cell is controlled by a PIN diode (SMP1320-079LF). The input image is converted into a binary image with the pixel size of 12 × 12, which is equivalently mapped to the corresponding unit cell of the metasurfaces. The near-field scanning device is used to measure the surface field and near field generated by the metasurfaces.

Each metasurface is controlled by a $L \times M \times N$ periodic time-varying sequence. The temporal modulation is contingent upon the source data matrices $D$ and trainable weight matrices $W$ in the time-varying sequence, which are systematically interlaced along the temporal-axis to govern each metasurface in the linear system. Therefore, the $i^{th}$ metasurface is modulated by $S^i = [D, W_1^i, D, W_2^i, D, W_l^i, ..., D, W_L^i]$, where $W_l^i$ corresponds to weight matrix at $l^{th}$ time partition. During the training of temporal neural network, each element of the weight matrices is updated based on the stochastic gradient descent algorithm.

### Information state sensitivity

By reversing the ON/OFF state of each unit cell of metasurface, we define the cross-entropy difference induced by such perturbation as the single-pixel information state sensitivity, and each unit is defined as a high or low sensitivity state according to the cross-entropy difference. For a time-varying sequence with the length $L$, each metasurface layer contains $L$ single-pixel information state sensitivities. Single-pixel information state sensitivity with the size of $M \times N$ is divided into high-sensitivity and low-sensitivity area, where the low-sensitivity area is subsequently used for other tasks at another frequency. The total information state sensitivity of each layer is obtained by the union of single-pixel information state sensitivities in each time interval. Specifically, only when the $L$ single-pixel information state sensitivities in the time-varying sequence are all in a low sensitivity state at $(x, y)$, the element at $(x, y)$ of the total information state sensitivity will be classified as a part of low sensitivity area. Alternatively, another approach involves summing $L$ single-pixel information state sensitivities after normalization, and then input the summation result into the threshold filter to derive the final outcome, whereas the area of summation results greater than the threshold corresponds to the high sensitivity area and the area of summation result below the threshold corresponds to low sensitivity area. By setting an appropriate threshold, the low-sensitivity area of total information state sensitivity is obtained and we broaden the length of time-varying sequences of this area, while the sequences of the other areas maintain the pre-defined configuration.

**Wake-sleep algorithm for asynchronous based Lagrangian optimization problem**

As mentioned above, the multi-frequency control problem can be redefined as a Lagrangian problem. Taking two parallel tasks as an example, the Lagrangian equation can be written as,

$$\mathcal{L}_{Lg}(\boldsymbol{W}^{\Delta f_2}; \boldsymbol{D}) = \mathcal{L}_{CE}(\boldsymbol{W}^{\Delta f_2}; \boldsymbol{D}) - \lambda \left( \mathbb{E}_{p_{data}(\boldsymbol{D})} \left[ \mathcal{L}_{CE}\left(\widehat{\boldsymbol{W}^{\Delta f_1}}; \boldsymbol{D}\right) \right] - \mathcal{L}_{CE}\left(\widehat{\boldsymbol{W}^{\Delta f_1}}, \boldsymbol{W}^{\Delta f_2}; \boldsymbol{D}\right) \right)$$

(6)

where $\mathcal{L}_{CE}(\boldsymbol{W}^{\Delta f_2}; \boldsymbol{D})$ is the cross-entropy function corresponding to network performance at modulation frequency $\Delta f_2$. Proposed constrain $\lambda \left( \mathbb{E}_{p_{data}(\boldsymbol{D})} \left[ \mathcal{L}_{CE}\left(\widehat{\boldsymbol{W}^{\Delta f_1}}; \boldsymbol{D}\right) \right] - \mathcal{L}_{CE}\left(\widehat{\boldsymbol{W}^{\Delta f_1}}, \boldsymbol{W}^{\Delta f_2}; \boldsymbol{D}\right) \right)$ in Eq. (6) is the pre-trained network cross entropy at modulation frequency $\Delta f_1$ averaged over the observational data, which is supposed to not increase during subsequent training. $\lambda$ is Lagrange multiplier, acting like balance factor. Based on a variant of the wake-sleep algorithm, we find an effective approach to optimize Eq. (6). Initially, all unit cells of the metasurfaces are modulated by frequency $\Delta f_1$, with corresponding time-varying sequence encompassing four trainable weight matrices. We firstly conduct preliminary training using temporal nonlinear network for an expression recognition task at $\Delta f_1$ on the observed dataset, yielding trained $\widehat{\boldsymbol{W}^{\Delta f_1}} = [\boldsymbol{W}_1^{\Delta f_1}, \boldsymbol{W}_2^{\Delta f_1}, \boldsymbol{W}_3^{\Delta f_1}, \boldsymbol{W}_4^{\Delta f_1}]$. Subsequently, utilizing the total information state sensitivity, low sensitivity regions of metasurface are defined as the $\Delta f_2$ frequency area, of which the length of temporal sequences is extended to 16, incorporating additional four trainable weight matrices $\boldsymbol{W}^{\Delta f_2} = [\boldsymbol{W}_5^{\Delta f_2}, \boldsymbol{W}_6^{\Delta f_2}, \boldsymbol{W}_7^{\Delta f_2}, \boldsymbol{W}_8^{\Delta f_2}]$. Then, the network is fine tunned only at $\Delta f_2$ frequency area for another gender recognition task, while keeping $\widehat{\boldsymbol{W}^{\Delta f_1}}$ frozen, i.e., updating the parameters $\boldsymbol{W}^{\Delta f_2}$ with $\widehat{\boldsymbol{W}^{\Delta f_1}}$ fixed. Through simple brute force search, we can achieve different classification tasks on two frequencies simultaneously with high classification accuracy. More harmonic manipulation can be achieved through stochastic gradient optimization.

**Reinforcement learning process for maze-solving**

We explore the possibility of using temporally-induced nonlinear system in a typical deep Q network (DQN). The Q network and target Q' network are assigned with identical trainable parameters at the beginning. During training process, the training samples are discrete, and all maze data are randomly shuffled and randomly extracted. Q network is updated in real-time, while the target Q' network undergoes intermittent updates to maintain the network robustness. After sufficient training, our reinforcement learning method achieves energy focusing in complex scenes through Algorithm 1, enabling the trained agent to provide correct direction policy for the vehicle in the maze in real time, which offers a novel option for intelligent driving[58]. The specific workflow is as follows.

**Algorithm 1:** deep Q-learning using spatiotemporal nonlinear systems

**Input:** data $x$, trainable weight matrix $W^{T_0} = [W_1^{T_0}, W_2^{T_0}, \ldots W_{127}^{T_0}, W_{128}^{T_0}]$,
**Output:** output field $U$

1. Initialize state buffers $D$ with capacity $K$
   Initialize action-value function Q (represented by temporal nonlinear system) with random weights $W^{T_0}$
2. **For** m = 1, M **do**
3.     Initialize sequence $s_1 = \{x_1\}$ and pre-processed sequenced $\phi_1 = \phi(s_1)$
4.     **For** t = 1, N **do**
5.         With probability $\varepsilon$ select a random action $a_t$, or select $a_t = \max_a Q^*(\phi(s_t), a; W^{T_0})$
6.         Execute $a_t$ in maze and obtain reward $r_t$ and updated maze data $x_{t+1}$
7.         Set $s_{t+1} = s_t$, $\phi_{t+1} = \phi(s_{t+1})$ and store $(\phi_t, a_t, r_t, \phi_{t+1})$ in $D$
8.         Randomly sample batch of $(\phi_j, a_j, r_j, \phi_{j+1})$ from $D$
9.         If terminal $\phi_{j+1}$, set $y_j = r_j$; Otherwise set $y_j = r_j + \gamma \max_{a'} Q^*(\phi_{j+1}, a'; W^{T_0})$
10.        Error $\delta_j = (Q(\phi_j, a_j; W^{T_0}) - y_j)^2$
11.        Update $W^{T_0} \leftarrow W^{T_0} - \alpha \cdot \delta_j \cdot \frac{\partial Q(\phi_j, a_j; W^{T_0})}{\partial W^{T_0}}$
12.        Every C steps reset $Q^* = Q$
13.     **End for**
14. **End for**

**Acknowledgements**
The work at Zhejiang University was sponsored by the National Natural Science Foundation of China (NNSFC) under Grant Nos. 62422514, 62471432, 62101485 and 61975176, the Key Research and Development Program of the Ministry of Science and Technology under Grant Nos. 2022YFA1404704, 2022YFA1405200, and 2022YFA1404902, the Key Research and Development Program of Zhejiang Province under Grant No. 2022C01036, and the Fundamental Research Funds for the Central Universities.

**Author contributions**
C.Q. and G. Y. conceived the idea. G.Y. performed the simulation and experiment. G.Y. and C.Q. wrote the paper. C.Q., and H.C. supervised the project.

**Competing interests**
The authors declare no competing interests.

**References**

1. Y. LeCun, Y. Bengio, & G. Hinton, Deep learning. *Nature* **521**, 436–444 (2015).

2. L. G. Wright, T. Onodera, M. M. Stein, T. Wang, D. T. Schachter, Z. Hu, & P. L. McMahon, Deep physical neural networks trained with backpropagation. *Nature* **601**, 549–555 (2022).

3. K. He, X. Zhang, S. Ren, & J. Sun, Deep residual learning for image recognition. In *Proc. IEEE Conference on Computer Vision and Pattern Recognition*. 770–778 (IEEE, 2016).

4. C. Qian, B. Zheng, Y. Shen, L. Jing, E. Li, L. Shen, & H. Chen, Deep-learning-enabled self-adaptive microwave cloak without human intervention. *Nat. Photon.* **14**, 383–390 (2020).

5. J. N. P. Martel, L. K. Mueller, S. J. Carey, P. Dudek, & G. Wetzstein, Neural sensors: learning pixel exposures for HDR imaging and video compressive sensing with programmable sensors. *IEEE Trans. Pattern Anal. Mach. Intell.* **42**, 1642–1653 (2020).

6. C. Qian, Z. Wang, H. Qian, T. Cai, B. Zheng, X. Lin, Y. Shen, I. Kaminer, E. Li, & H. Chen, Dynamic recognition and mirage using neuro-metamaterials. *Nat. Commun.* **13**, 2694 (2022).

7. Z. Huang, W. Shi, S. Wu, Y. Wang, S. Yang, & H. Chen, Pre-sensor computing with compact multilayer optical neural network. *Sci. Adv.* **10**, eado8516 (2024).

8. C. Qian, X. Lin, X. Lin, J. Xu, Y. Sun, E. Li, B. Zhang, & H. Chen, Performing optical logic operations by a diffractive neural network. *Light Sci. Appl.* **9**, 59 (2020).

9. T. Yan, J. Wu, T. Zhou, H. Xie, F. Xu, J. Fan, L. Fang, X. Lin, & Q. Dai, Fourier-space Diffractive Deep Neural Network. *Phys. Rev. Lett.* **123**, 023901 (2019).

10. X. Lin, Y. Rivenson, N. T. Yardimci, M. Veli, Y. Luo, M. Jarrahi, & A. Ozcan, All-optical machine learning using diffractive deep neural networks. *Science* **361**, 1004–1008 (2018).

11. G. M. Marega, H. G. Ji, Z. Wang, G. Pasquale, M. Tripathi, A. Radenovic, & A. Kis, A large-scale integrated vector–matrix multiplication processor based on monolayer molybdenum disulfide memories. *Nat. Electron.* **6**, 991–998 (2023).

12. Y. Zuo, B. Li, Y. Zhao, Y. Jiang, Y. C. Chen, P. Chen, G. B. Jo, J. Liu, & S. Du, All-optical neural network with nonlinear activation functions. *Optica* **6**, 1132–1137 (2019).

13. R. A. Heinz, J. O. Artman, & S. H. Lee, Matrix multiplication by optical methods. *Appl. Opt.* **9**, 2161–2168 (1970).

14. C. Cuppini, L. Shams, E. Magosso, & M. Ursino, A biologically inspired neurocomputational model for audiovisual integration and causal inference. *Eur. J. Neurosci.* **46**, 2481–2498 (2017).

15. F. Ashtiani, A. J. Geers, & F. Aflatouni, An on-chip photonic deep neural network for image classification. *Nature* **606**, 501–506 (2022).

16. C.Y. Shen, J. Li, T. Gan, Y. Li, M. Jarrahi, & A. Ozcan, All-optical phase conjugation using diffractive wavefront processing. *Nat. Commun.* **15**, 4989 (2024).

17. C.Y. Shen, J. Li, Y. Li, T. Gan, L. Bai, M. Jarrahi, & A. Ozcan, Multiplane quantitative phase imaging using a wavelength-multiplexed diffractive optical processor. *Adv. Photonics* **6**, 056003 (2024).

18. Z. Fan, C. Qian, Y. Jia, Z. Wang, Y. Ding, D. Wang, L. Tian, E. Li, T. Cai, B. Zheng, I. Kaminer, & H. Chen, Homeostatic neuro-metasurfaces for dynamic wireless channel management. *Sci. Adv.* **8**, eabn7905 (2022).

19. B. Wu, W. Zhang, H. Zhou, J. Dong, D. Huang, P. K. A. Wai, & X. Zhang, Chip-to-chip optical multimode communication with universal mode processors. *PhotoniX* **4**, 37 (2023).

20. Z. Xue, T. Zhou, Z. Xu, S. Yu, Q. Dai, & L. Fang, Fully forward mode training for optical neural networks. *Nature* **632**, 280–286 (2024).

21. D. Zhang, D. Xu, Y. Li, Y. Luo, J. Hu, J. Zhou, Y. Zhang, B. Zhou, P. Wang, X. Li, B. Bai, H. Ren, L. Wang, A. Zhang, M. Jarrahi, Y. Huang, A. Ozcan, & X. Duan, Broadband nonlinear modulation of incoherent light using a transparent optoelectronic neuron array. *Nat. Commun.* **15**, 2433 (2024).

22. C. Qian, Z. Wang, H. Qian, T. Cai, B. Zheng, X. Lin, Y. Shen, I. Kaminer, E. Li, & H. Chen, Dynamic recognition and mirage using neuro-metamaterials. *Nat. Commun.* **13**, 2694 (2022).

23. J. Yu, X. Yang, G. Gao, Y. Xiong, Y. Wang, J. Han, Y. Chen, H. Zhang, Q. Sun, & Z. Wang, Bioinspired mechano-photonic artificial synapse based on graphene/$MoS_2$ heterostructure. *Sci. Adv.* **7**, eabd9117 (2021).

24. G. Giusfredi, S. Bartalini, S. Borri, P. Cancio, I. Galli, D. Mazzotti, & P. D. Natale, Saturated-Absorption Cavity Ring-Down Spectroscopy, *Phys. Rev. Lett.* **104**, 110801 (2010).

25. A. Baas, J. Karr, H. Eleuch, & E. Giacobino, Optical bistability in semiconductor microcavities. *Phys. Rev. A* **69**, 023809 (2004).

26. F. Xia, K. Kim, Y. Eliezer, S. Han, L. Shaughnessy, S. Gigan, & H. Cao, Nonlinear optical encoding enabled by recurrent linear scattering. *Nat. Photon.* **18**, 1067–1075 (2024).

27. M. Yildirim, N. U. Dinc, I. Oguz, D. Psaltis, & C. Moser, Nonlinear processing with linear optics. *Nat. Photon.* **18**, 1076–1082 (2024).

28. C. C. Wanjura & F. Marquardt, Fully nonlinear neuromorphic computing with linear wave scattering. *Nat. Phys.* **20**, 1434–1440 (2024).

29. W. Nie, Optical nonlinearity-phenomena, applications and materials. *Adv. Mater.* **5**, 520–545 (1993).

30. Y. Li, J. Li, & A. Ozcan, Nonlinear encoding in diffractive information processing using linear optical materials. *Light Sci. Appl.* **13**, 173 (2024).

31. C. Qian, Y. Jia, Z. Wang, J. Chen, P. Lin, X. Zhu, E. Li & H. Chen, Autonomous aeroamphibious invisibility cloak with stochastic-evolution learning. *Adv. Photonics* **6**, 016001 (2024).

32. J. Sisler, P. Thureja, M. Y. Grajower, R. Sokhoyan, I. Huang, & H. A. Atwater, Electrically tunable space-time metasurfaces at optical frequencies. *Nat. Nanotechnol.* **19**, 1491–1498 (2024).

33. C. Liu, Q. Ma, Z. Luo, Q. Hong, Q. Xiao, H. Zhang, L. Miao, W. Yu, Q. Cheng, L. Li, & T. Cui, A programmable diffractive deep neural network based on a digital-coding metasurface array. *Nat. Electron.* **5**, 113–122 (2022).

34. Y. Jia, C. Qian, Z. Fan, T. Cai, E. Li, & H. Chen, A knowledge-inherited learning for intelligent metasurface design and assembly. *Light Sci. Appl.* **12**, 82 (2023)

35. J. W. Liu, G. G. Liu, & B. Zhang, Three-dimensional topological photonic crystals. *Prog. Electromagn. Res.* **181**, 99–112 (2024).

36. Z. Fan, Y. Jia, H. Chen, & C. Qian, Spatial multiplexing encryption with cascaded metasurfaces. *J. Opt.* **25**, 125105 (2023).

37. D. Bouchet, S. Rotter, & A. P. Mosk, Maximum information states for coherent scattering measurements. *Nat. Phys.* **17**, 564–568 (2021).

38. J. W. Goodman, Introduction to Fourier Optics (Roberts and Company, 2005).

39. A. Tennant & B. Chambers, Time-switched array analysis of phase-switched screens. *IEEE Trans. Antenn. Propag.* **57**, 808–812 (2009).

40. Z. L. Zhang & M. R. Sabuncu, Generalized cross entropy loss for training deep neural networks with noisy labels. In *Proceedings of the 32nd International Conference on Neural Information Processing Systems*. 8792–8802 (ACM, Montréal, 2018).
41. Y. Jia, H. Lu, Z. Fan, B. Wu, F. Qu, M. J. Zhao, C. Qian, & H. Chen, High-efficiency Transmissive tunable metasurfaces for binary cascaded diffractive layers. *IEEE Trans. Antenn. Propag*. **72**, 5 (2024).
42. A. M. Shaltout, V. M. Shalaey, & M. L. Brongersma, Spatiotemporal light control with active metasurfaces. *Science* **364**, eaat3100 (2019).
43. Z. Zhang, P. Luo, C. C. Loy, & X. Tang, Facial landmark detection by deep multi-task learning. *Proc. Eur. Conf. Comput. Vis,* 94–108 (2014).
44. O. Kulce, D. Mengu, Y. Rivenson, & A. Ozcan, All-optical information-processing capacity of diffractive surfaces. *Light Sci. Appl.* **10**, 25 (2021).
45. S. Zagoruyko & N. Komodakis, Paying more attention to attention: improving the performance of convolutional neural networks via attention transfer. In *International Conference on Learning Representations* (ICLR, 2017).
46. A. D. Brink & N. E. Pendock, Minimum cross-entropy threshold selection. *Pattern Recognit.* **29**, 179–188 (1996).
47. J. Feldmann, N. Youngblood, M. Karpov, H. Gehring, X. Li, M. Stappers, M. L. Gallo, X. Fu, A. Lukashchuk, A. S. Raja, J. Liu, C. D. Wright, A. Sebastian, T. J. Kippenberg, W. H. P. Pernice, & H. Bhaskaran, Parallel convolutional processing using an integrated photonic tensor core. *Nature* **589**, 52–58 (2021).
48. K. Kim, S. Bittner, Y. Zeng, S. Guazzotti, O. Hess, Q. Wang, & H. Cao, Massively parallel ultrafast random bit generation with a chip-scale laser. *Science* **371**, 948–952 (2021).
49. S. Wang, M. Chen, J. Ke, Q. Cheng, & T. Cui, Asynchronous space-time-coding digital metasurface. *Adv. Sci.* **9**, 2200106 (2022).
50. J. Hüpfl, F. Russo, L. M. Rachbauer, D. Bouchet, J. Lu, U. Kuhl, & S. Rotter, Continuity equation for the flow of Fisher information in wave scattering. *Nat. Phys.* **20**, 1294–1299 (2024).
51. G. E. Hinton, P. Dayan, B. J. Frey, & R. M. Neal, The "wake-sleep" algorithm for unsupervised neural networks. *Science* **268**, 1158–1161 (1995).
52. Z. Fan, C. Qian, Y. Jia, Y. Feng, H. Qian, E. Li, R. Fleury, & H. Chen, Holographic multiplexing metasurface with twisted diffractive neural network. *Nat. Commun.* **15**, 9416 (2024).
53. F. Paredes-Vallés, J. J. Hagenaars, J. Dupeyroux, S. Stroobants, Y. Xu, & G. C. H. E. Decroon, Fully neuromorphic vision and control for autonomous drone flight. *Sci. Robot.* **9**, eadi0591 (2024).
54. G. He, C. Qian, Y. Jia, Z. Fan, H. Wang, & H. Chen, Twisted metasurfaces for on-demand focusing localization. *Adv. Opt. Mater.* **13**, 2570041 (2024).
55. M. Gu, Q. Zhang, & S. Lamon, Nanomaterials for optical data storage. *Nat. Rev*. *Mater.* **1**, 16070 (2016).
56. J. Deng, Z. Li, J. Li, Z. Zhou, F. Gao, C. Qiu, & B. Yan, Metasurface-assisted optical encryption carrying camouflaged information. *Adv. Opt. Mater.* **10**, 2200949 (2022).
57. L. Wang, X. Mao, A. Wang, Y. Wang, Z. Gao, S. Li, & L. Yan, Scheme of coherent optical chaos communication. *Opt. Lett.* **45**, 4762–4765 (2020).

58. Y. Song, A. Romero, M. Müller, V. Koltun, & D. Scaramuzza, Reaching the limit in autonomous racing: optimal control versus reinforcement learning. *Sci. Robot.* **8**, eadg1462 (2023).

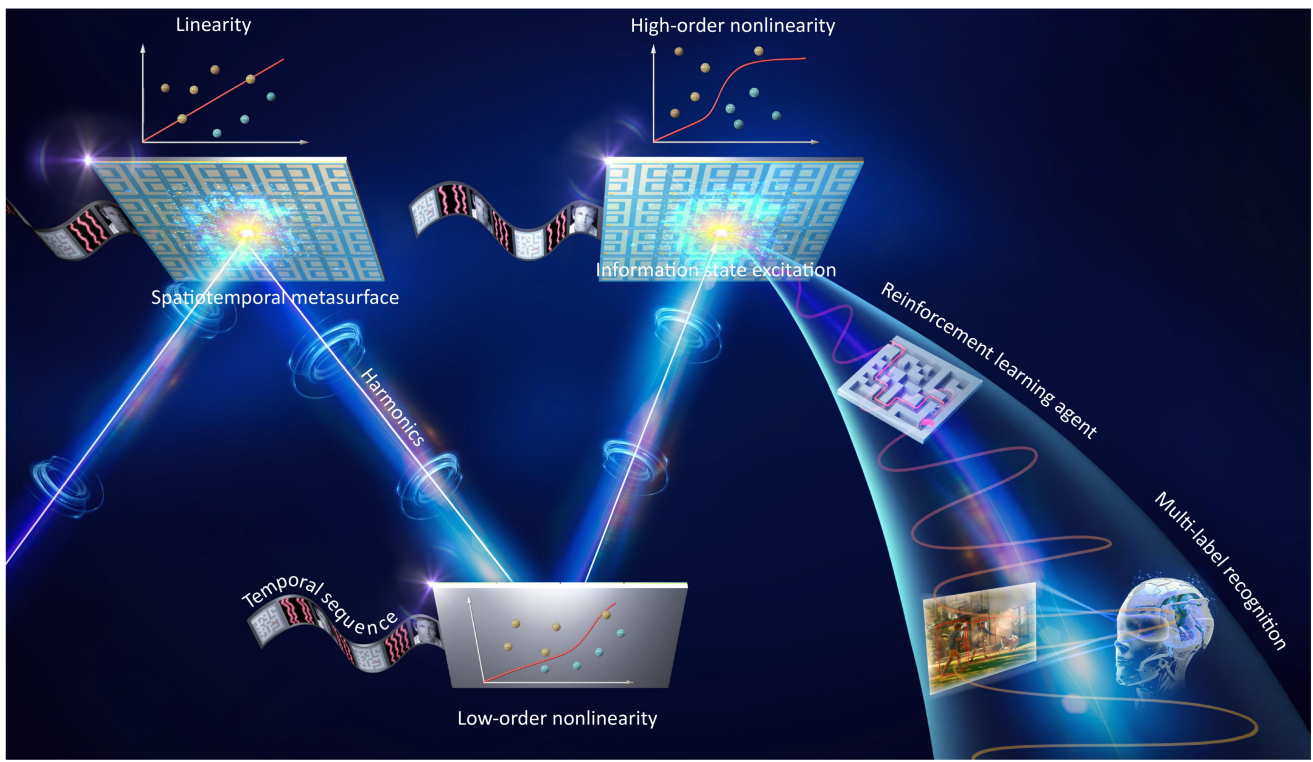

**Fig. 1 | Parallel nonlinear metasurface processor with temporal encoding.** By utilizing the temporal encoding strategy, both data and weight matrices are concurrently mapped to each metasurface without necessitating additional encoding functions. For a given incidence field, the data information propagates between the metasurfaces via linear diffraction, while high-order states of the data (i.e., nonlinearity) are excited during multiple scattering. The proposed nonlinear processor is thus capable of implementing trainable neural network and efficiently processing complex high-dimensional information. Upon sustained modulation of multiple metasurface layers, the linear system response undergoes a transformation into the nonlinear response. By appropriately configuring the weight matrices, the temporal nonlinear network can be effectively utilized into various deep application scenarios, such as maze-solving and object recognition.

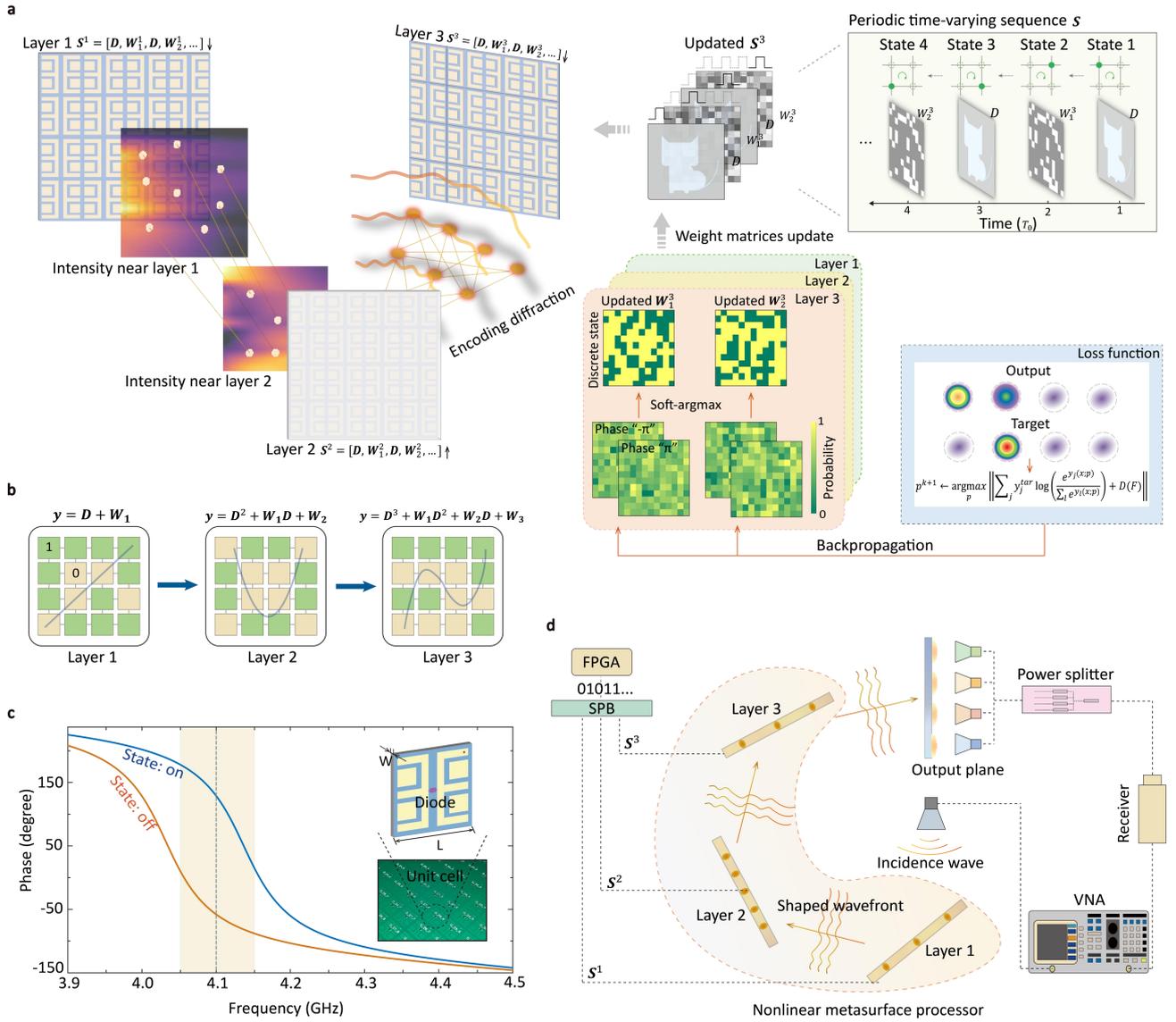

**Fig. 2 | Mechanism for temporal nonlinear neural network through distributed spatiotemporal metasurfaces. a**, Schematic diagram of the temporal neural network architecture. The temporal sequence $S$ is composed of multiple individual information states that cycle periodically to control the metasurfaces. **b**, Nonlinear response evolution across the metasurface layers. In the second and third layers, employing the time encoding method in conjunction with data repetition strategy, high-order terms of data are excited through multiple diffraction, allowing the output to exhibit flexible nonlinearity. Moreover, this encoding strategy fosters the generation of higher-order nonlinearity as the number of the metasurfaces increases. **c**, Reflection response of the designed spatiotemporal metasurfaces. The unit cell exhibits a phase difference of at least 180 degrees in the frequency range of 4.05 GHz to 4.15 GHz. **d**, Experimental setup of nonlinear metasurface processor.

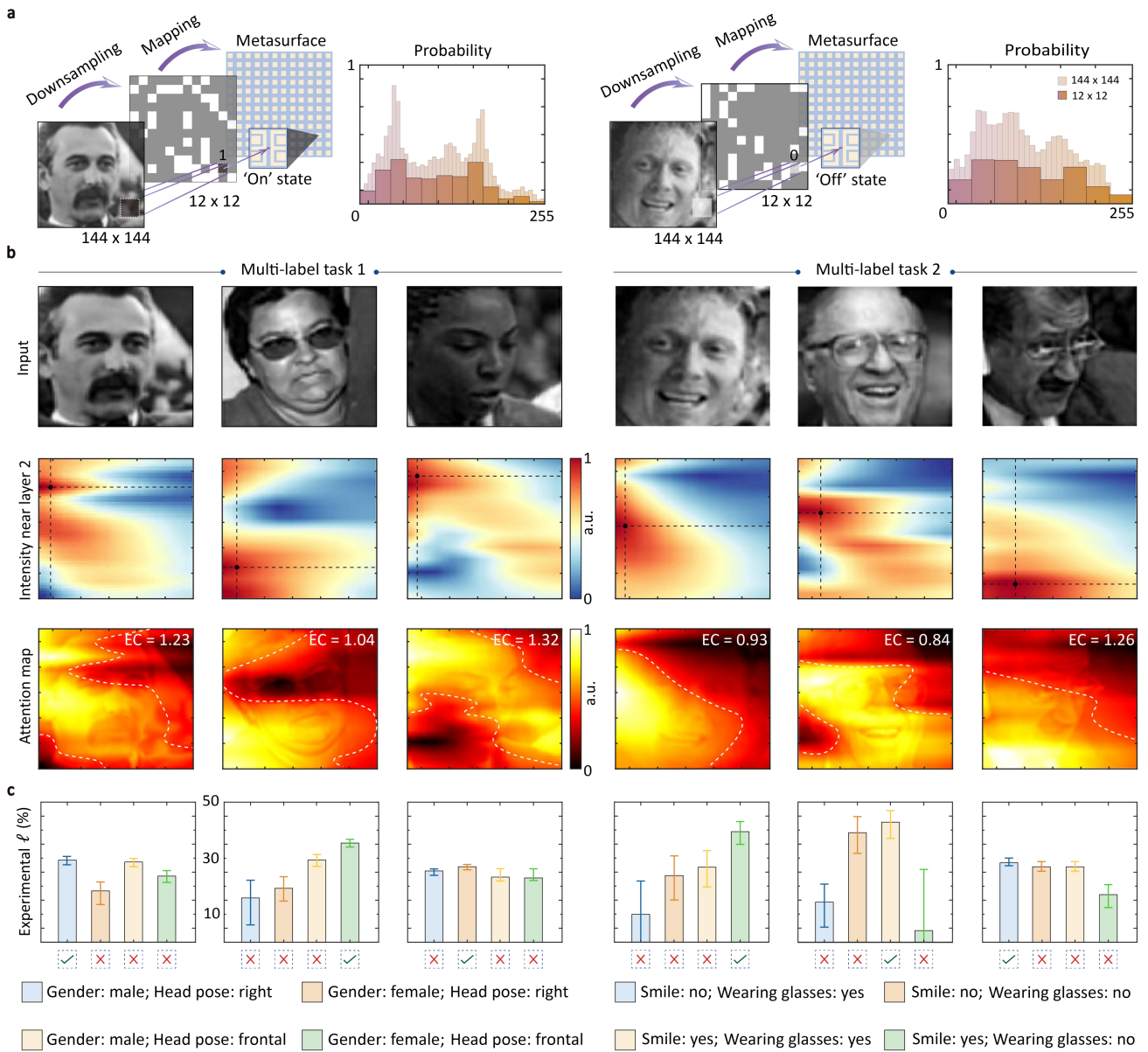

**Fig. 3 | Multi-label recognition task with nonlinear metasurface processor. a**, Data processing and mapping. The original image is squeezed into 12 × 12 down-sampled binary image and mapped into the ON/OFF state distribution of metasurface. It can be observed that the probability distribution of the squeeze image retains the main features of the original data from the histogram. **b**, Intensity distribution near layer 2 corresponding to different input data. The attention map derived from the intensity distribution highlights the area of the image features that affects the classification accuracy most. **c**, Performance analysis of facial recognition. The bars denote the mean intensity measurement of four detectors and the error bars represent the standard deviation of intensity. EC, entropy coefficient.

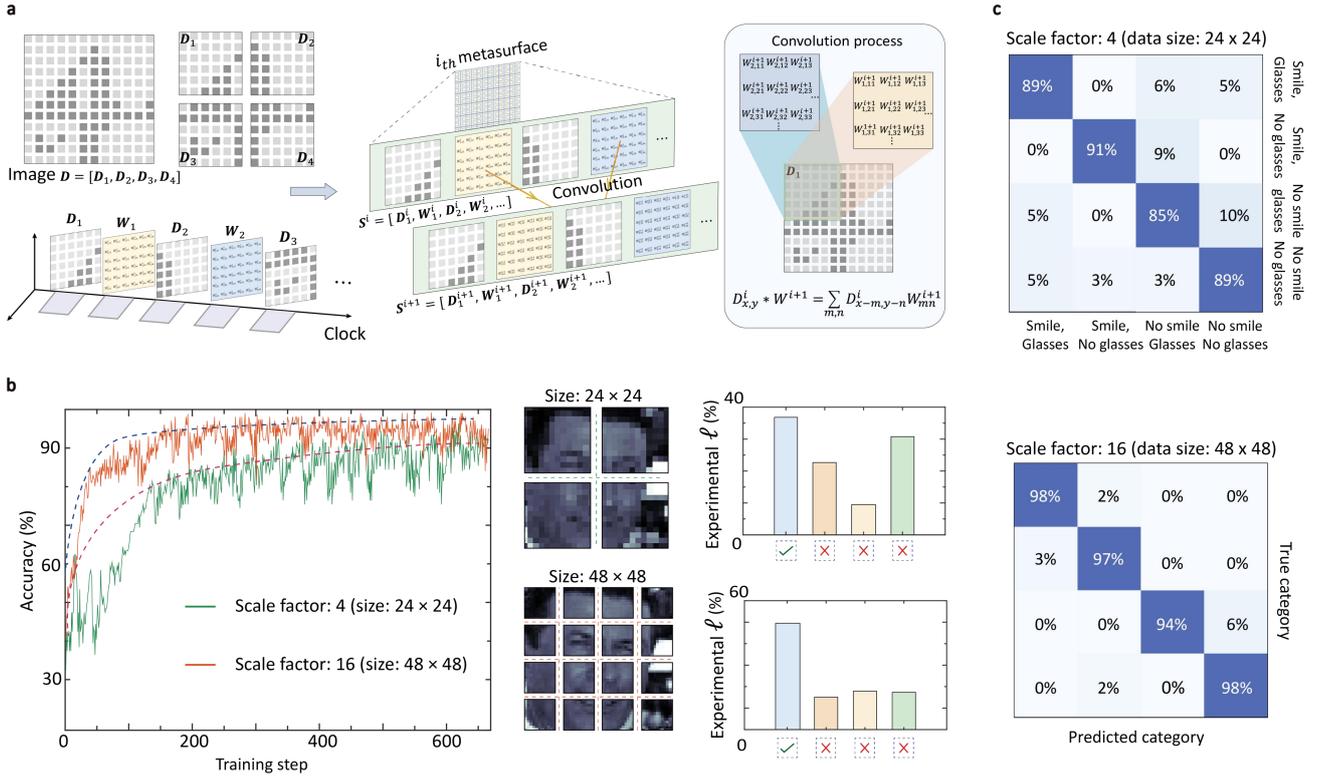

**Fig. 4 | Demonstration of residual convolutional operation and scalability in nonlinear metasurface processor. a**, Convolution process in our temporal nonlinear framework. The input data $D$ is divided into $[D_1, D_2, D_3, D_4]$ and cross-arranged with weight matrices into time-varying sequence. During diffraction process, multiple weight matrices at $(i+1)_{th}$ layer are convolved with the image $D$ at $i_{th}$ layer to generate compressed information feature map, which undergoes successive compression in subsequent metasurface layers, thereby enabling the nonlinear network to capture high-dimensional image features. Simultaneously, as each metasurface layer receives the raw image data $D$, the input for each layer comprises both the compressed information feature map and the original image information, thus establishing a form of residual convolution. Such improved residual convolution mechanism mitigates the risk of the network neglecting the overall features of the images, ultimately contributing to enhanced network performance. **b**, Evolution of the test accuracy during training for the scale factor 4 and 16. Recognition results of an input image with different data size are visualized. **c**, Confusion matrix corresponding to data size of 24 × 24 and 48 × 48.

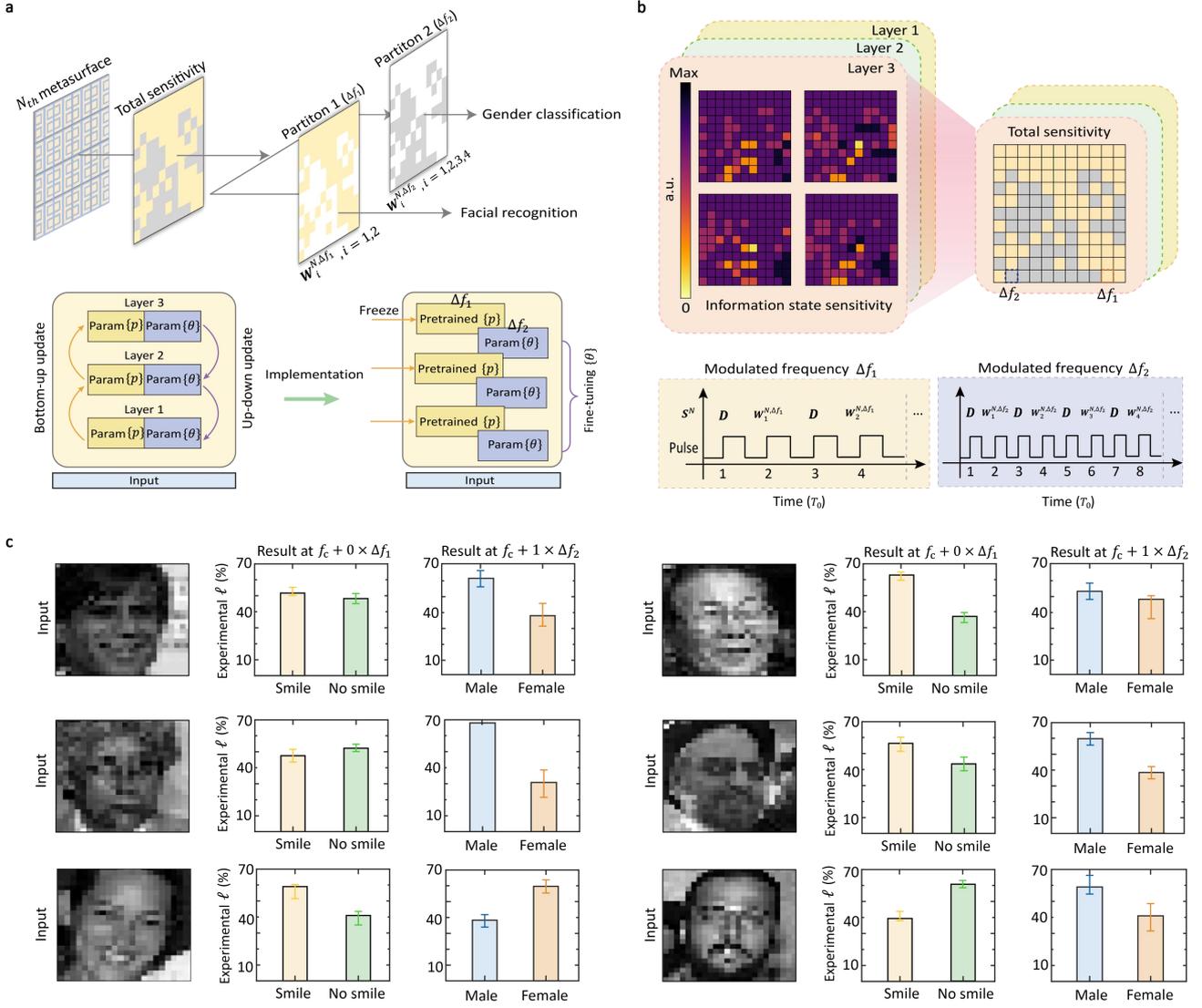

**Fig. 5 | Nonlinear multi-task demonstration with asynchronous modulation**. **a**, Gender classification and expression recognition synchronously with asynchronous modulation using modified "wake-sleep" algorithm. The object in Eq. (6) can be optimized by a variant of "wake-sleep" algorithm. Instead of alternately optimizing the weight parameters in the $\Delta f_1$ frequency partition and weight parameters in the $\Delta f_2$ frequency partition, pretrained results at modulation frequency $\Delta f_1$ is utilized as correction term for Eq. (6). Upon finishing preliminary network training at frequency $\Delta f_1$, we perform perturbations to the local areas of metasurface assigned to modulation frequency $\Delta f_2$ by randomly altering the states of the unit cells of metasurface and update the weight matrices according to Eq. (6). Consequently, the network at $\Delta f_2$ frequency area is optimized without performance degradation at modulation frequency $\Delta f_1$. **b**, Information sensitivity of each metasurface layer. Employing the information state sensitivity, each metasurface is delineated into two distinct frequency regions, governed by different time-varying sequence. **c**, Recognition result at the main frequency and harmonic.

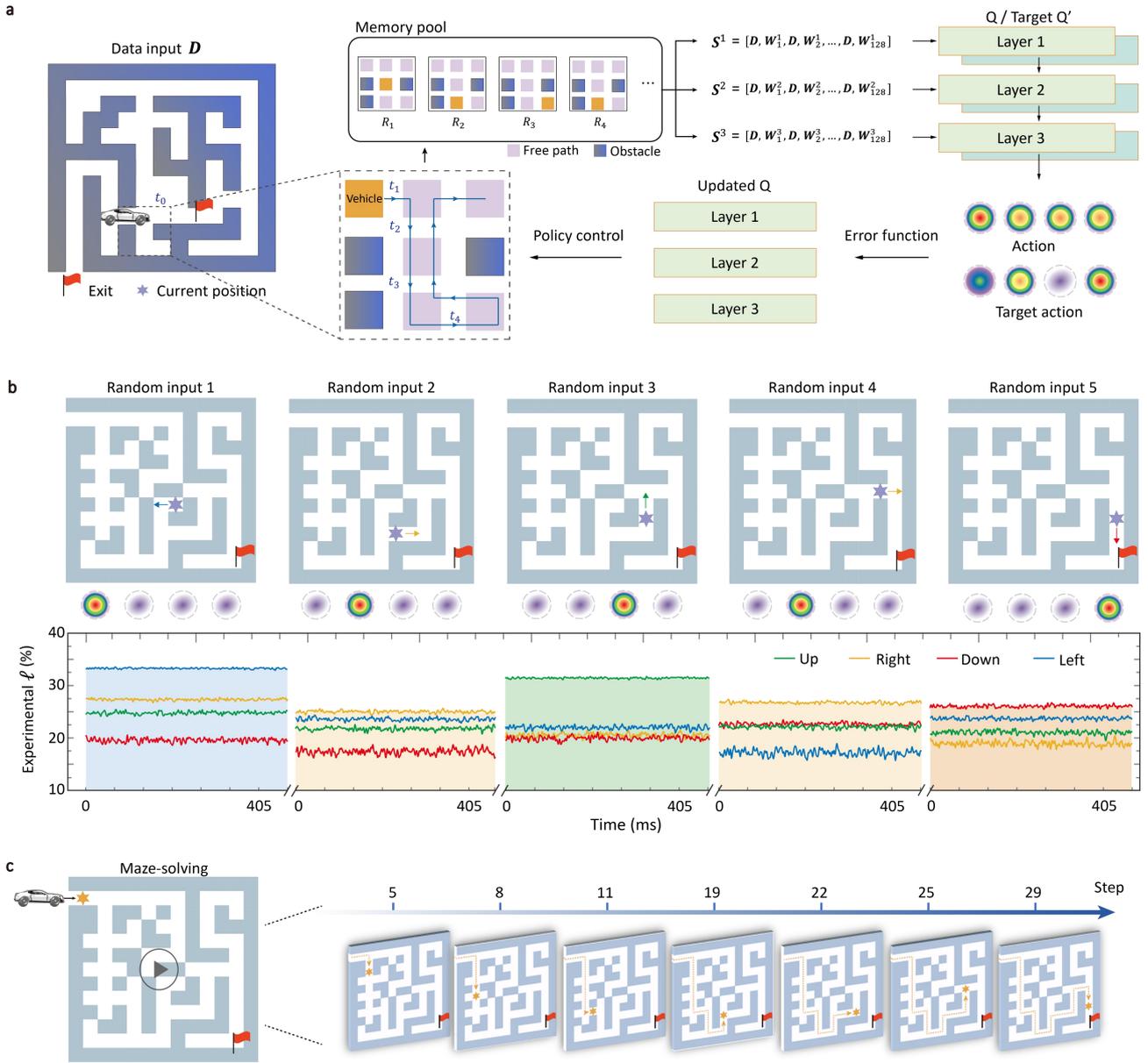

**Fig. 6 | Maze-solving with nonlinear metasurface agent. a**, Training process of maze-solving using temporal neural network. Our network optimizes actions through the instant reward-punishment mechanism, which can be trained through real-time iterative exploration. In the training process, the metasurface agent drives the vehicle to continuously explore new scenes and updates the network parameters based on continuous reward feedback. Furthermore, this unique training mechanism endows the metasurface agent with dynamic memory capability. The agent retains scene information explored by vehicle and make improved strategy based on previous explorations during subsequent scene exploration, which allows the vehicle to retrace from erroneous exploration paths and recommence exploring again. **b**, Different policy response of nonlinear agent at random location. The strongest variance of intensity aligns with true label. **c**, Complete maze solving process from the top left to the bottom right corner.